\documentclass[11pt,twoside]{article}
\usepackage{macroaaa-eng}
\usepackage{psfig}
\usepackage{graphicx}

\usepackage[T1]{fontenc} % Computer Modern (CM) fonts

\usepackage{latexsym}
\usepackage{verbatim}

\begin{document}

\vskip 1.0cm
\markboth{Botti et al.}{BLR-size and black hole mass in high-z AGN}
\pagestyle{myheadings}

\vspace*{0.5cm}
\title{Broad-line region size and black hole mass in high-z AGN}

\author{Ismael Botti, Paulina Lira and Jos\'e Maza}
\affil{Departamento de Astronom\'ia, Universidad de Chile, Casilla 36-D, Santiago, Chile.}
\vspace{0.2cm}
\author{Hagai Netzer, Shai Kaspi and Dan Maoz}
\affil{School of Physics and Astronomy and the Wise Observatory, Tel-Aviv University, Tel-Aviv 69978, Israel}

%%%%%% ABSTRACT %%%%%%%%%%%%%%%%%%%%%%%%%%%%%%%%%%%%%%%%%%%%%%%%%%%%%%%%%%%%%%%%%%%%%%%%%%%%%%%%%%%%%%%%%%%%%%%%

\begin{abstract} 
\small
In this contribution we briefly review the reverberation mapping technique and its results for low and intermediate luminosity AGNs. Then we present a monitoring campaign of high-luminosity high-redshift quasars which will extend these results by two orders of magnitude, probing the broad-line region size and black hole (BH) mass of luminous AGN at redshift $\sim$2-3.
\end{abstract}
%%%%%% INTRODUCTION %%%%%%%%%%%%%%%%%%%%%%%%%%%%%%%%%%%%%%%%%%%%%%%%%%%%%%%%%%%%%%%%%%%%%%%%%%%%%%%%%%%%%%%%%%%%
\section{Introduction}
The Broad-Line Region (BLR) is one of the main components of an unobscured AGN (ie, those where we have an unimpaired view torwards the nucleus), and is charactized by very broad emission lines in the UV/optical spectra (Figure 1) with typical widths of several thounsands km/sec due to the proximity to the central BH. One of the features of this region is that the observable UV/optical continuum flux variations are closely related to variations of the emission-lines fluxes. Combining the correlation between continuum and line fluxes variations with the widths of the lines, the mass of the central black hole can be estimated through the reverberation mapping technique (see \S2).\\
In these days, when it is widely accepted that all massive galaxies harbor a black hole in their centers, reverberation has allowed real physical comparison between active and dormants BHs in the local universe, but so far has not probed the high end of the AGN luminosity range.\\
In the next section we describe briefly the basic theory and results from previous reverberation mapping studies leading to the motivation for a high-luminosity high-redshift quasars monitoring campaign that we are performing since the beggining of 2005. Our main aim is to extend these results by two orders of magnitude, giving the firsts measurement of the largest BHs, and thus extending our knowledge of the physics of AGN and their hosts into an epoch crucial for the understanding of galaxy evolution.
%%%%%% REVERBERATION MAPPING BASICS %%%%%%%%%%%%%%%%%%%%%%%%%%%%%%%%%%%%%%%%%%%%%%%%%%%%%%%%%%%%%%%%%%%%%%%%%%%%
\section {Reverberation Mapping Basics}
Reverberation mapping is a technique based on the response of the BLR gas to changes in the central continuum source. It is based on three basic assumptions:\newline
\begin{enumerate}
	\item \emph{The ionizing continuum originates in a single central source}.\medskip
	\\This holds because the BLR size is $\sim$2 orders of magnitude larger than the continuum source.
	\item \emph{The light travel time across the BLR is the most important time scale.} \medskip
	\\Which means that the clouds response is practically ``instantaneous''. Indeed, the recombination time is related with the electron density $n_{e}$ by the relation $\tau_{rec} \approx 0.1(10^{10}\mbox{ cm}^{-3}/n_e)\mbox{ hr}\approx 0.1\mbox{ hr}$, which is much less than the typical light travel time across the BLR.
	\item \emph{There is a simple but not necessary linear relationship between the observed continuum and the ionizing continuum (which gives rise to the emission lines).}
\end{enumerate}
From these points one can construct a simple 1D linear model given by 
\begin{equation}
	 L(t) = \int \Psi(\tau) C(t-\tau) d\tau
\end{equation}
where $L(t)$ and $C(t)$ are the line and continuum light curves respectively. $\Psi(\tau)$ is the \emph{transfer function} (TF) which holds the information about the geometry and kinematics of the BLR gas. Due to technical limitations, the TF is collapsed to one single parameter: the time lag $\tau$ between both continuum and line light curve variations which is obtained through a cross-correlation analysis between them (Peterson 2001). This lag $\tau$ is used to directly measure the BLR size from
\begin{equation}
	R_{BLR}= c\tau
\end{equation}
(Peterson et al., 2004, Kaspi et al., 2005). Assuming a gravitationally bound system (proven to be correct for those objects with measurements from several lines, where the observed anticorrelation between the distance $r$ to the BH and the line Doppler widths is consistent with virialized motions of the BLR gas) and measuring the BLR line widths $\sigma$, it is possible to infer the mass of the central BH through
\begin{equation}
	M_{BH} = \frac{f\ \sigma^{2}\ R_{BLR}}{G}
\end{equation}
where $f$ depends on the geometry of the BLR and it is expected to be $\sim$1. 
%%%%%% REVERBERATION RESULTS %%%%%%%%%%%%%%%%%%%%%%%%%%%%%%%%%%%%%%%%%%%%%%%%%%%%%%%%%%%%%%%%%%%%%%%%%%%%%%%%%%%
\section {Reverberation Results from low and intermidiate luminosity AGN}
Many important results have been found from reverberation mapping studies, such as the stratification of the BLR and black hole masses. Also, an empirical relation between the BLR size and the AGN luminosity has been determined, which shows that \emph{the size of the broad-line region scales with luminosity}. This is not far from the $R_{BLR} \propto L^{0.5}$ relation expected if the BLR in all AGN have similar $n_{e}$ value and ionization conditions. Kaspi et al. (2005) found that the scaling relation has the form 
\begin{equation}\label{RL}
	R_{BLR} \simeq 0.4 \left[\frac{\lambda L_{\lambda}(5100\mbox{ \AA})}{10^{46}\mbox{ erg/sec}}\right]^{0.6\pm0.1}\mbox{ pc}
\end{equation}
(as it is shown in Figure 1b). Preliminary results by Bentz et al (2006) suggest a smaller slope of about 0.52. Using $R_{\mbox{\tiny BLR}}-L$ relation is thus possible to estimate the BLR radius using $L$ alone, hence making BH mass estimation accessible for a large number of AGN. Such results cannot be directly applied to high-$z$, high-$L$ sources that contain the most massive BHs, since measuring their BLR size requires an extrapolation by up to two orders of magnitude in luminosity in eq. (\ref{RL}),
which would be subject of a large uncertainty in the measured $R_{BLR}$ (and then in the BH mass). This is the starting point of our quasars monitoring program.
\begin{figure}[t!]
\label{RvsL}
\begin{center}
\hbox{
	\includegraphics[width=6.9cm,height=5.8cm]{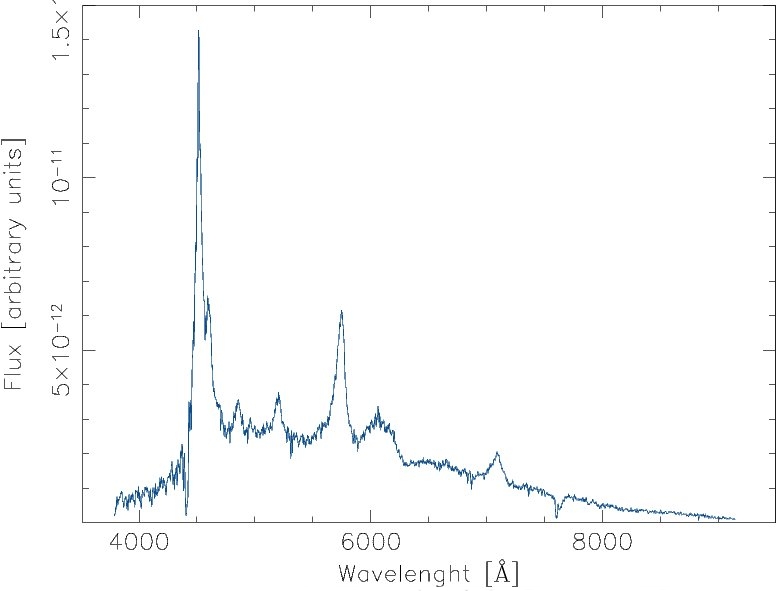}
	\includegraphics[width=6.4cm,height=5.8cm]{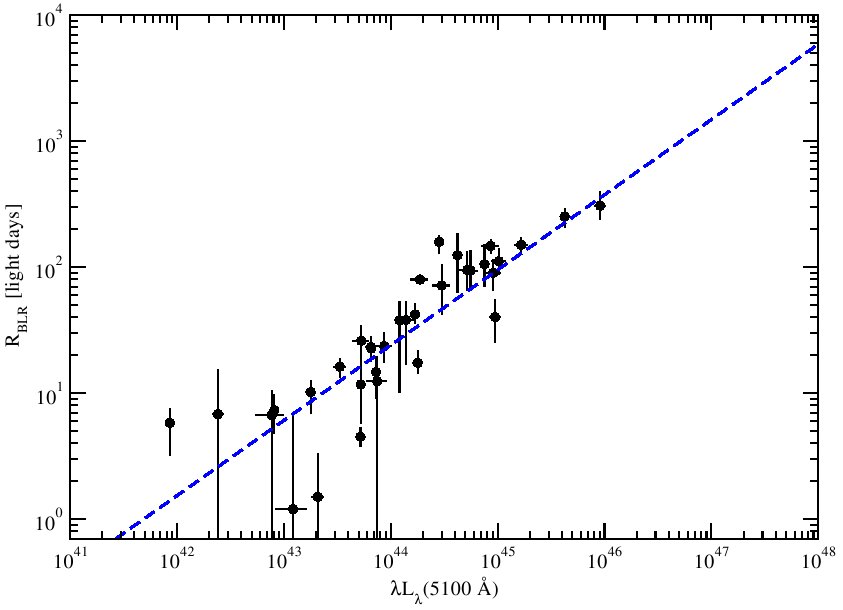}
        }
\caption{ \small {\it Left: \rm Optical spectrum of quasar CT260. Several UV broad-lines (Ly$\alpha$, Si\textsc{iv}, C\textsc{iv}, C\textsc{iii}) can be seen in the optical window indicating the high redshift of the object ($z=3$).} 
{\it Right: \rm Size-Luminosity relation for AGN with reverberation mapping from Kaspi et al. (2005). All our sources fall above the most luminous object in this diagram.} }
\end{center}
\end{figure}
%%%%%% QUASARS MONITORING %%%%%%%%%%%%%%%%%%%%%%%%%%%%%%%%%%%%%%%%%%%%%%%%%%%%%%%%%%%%%%%%%%%%%%%%%%%%%%%%%%%%%%
\section{Expanding the Luminosity Range: \textit{Quasar Monitoring Campaing}}
The few attempts to extend the luminosity range of AGN with reverberation mapping results to high-luminosity AGNs have failed for two reasons: 
1. Most high-$L$ sources are very slow variables with very low amplitude (< 20\%). 2. The line luminosity amplitude is even smaller than the continuum amplitude because the emission line response is averaged over the very large ($\sim 1$ pc) BLR geometry. %This requires the ability to measure line fluxes to an accuracy of about 2\%.\\
Since 2005 we are undertaking a novel strategy that uses optical broad band imaging to trigger spectroscopic follow up on a time scale of about one year. The idea is to identify high-$L$ sources with large enough continuum variations that will enable accurate and meaningful measurement of emission line variations. Only a small fraction of high-$L$ sources show such variations at any given time, hence, a large number of sources must be constantly monitored to identify the onset of such events. Given large enough continuum variations of a certain AGN, the event must be followed spectroscopically. The expected BLR size suggests that the spectroscopy phase should continue for at least four years.
\newline
Our sample consists of 56 quasars which are the brighest from the Calan-Tololo, 2dF and SDSS catalogues spread over all RAs, with $\lambda L_{\lambda}(5100)\ge10^{46.5}$ erg/sec, redshifts that range from 2.3 to 3.3 and mean R magnitude about $\sim$18.2. Until today, black hole mass estimations using reverberation mapping analysis have been mainly based in the variability of the H$\beta$ ($\lambda$4861\AA) emission line which lies in the optical. In our optical monitoring program we are looking for the C\textsc{iv} ($\lambda$1549\AA) emission line variability which, because of the the high redshifts of the ojects, falls into the optical window. Reverberation mapping based in the H$\beta$ line can be made for these objects going to the IR.
\subsection{Imaging monitoring.}
Imaging is obtained with the 0.9m Telescope at Cerro Tololo Interamerican Observatory (CTIO) every month for a sub-sample of quasars to monitor their continuum emission in the R band. Monthly datasets allow us to have from 4 to 5 light curves points in average per object in a year, enough to detect the onset of a high variation in their fluxes.\\
We use differential photometry respect 8 to 13 field stars which allows a photometry independent of the weather conditions, since the object and the field stars are observed simultaneously. These field stars are selected near to the quasar in order to appear in all the observations and have a magnitude cut-off of order of the quasar magnitude. To apply differential photometry we average the field star magnitudes in order to minimize the random variability of each of them, then
\begin{equation}
	\left\langle\ S\ \right\rangle(t) = \frac{1}{N}\sum_{i=1}^{N} m_{i}(t)
\end{equation}
is the average field star magnitude. The \emph{rms} of each field star light curve is also checked periodically to look for intrinsically variable objects. This procedure gives us an average magnitude almost constant within the error bars ($\left\langle S\right\rangle\sim3\%$ variable with a typical value of $\sim2\%$). The quasar Differential Light Curve (DLC) is given by
\begin{equation}
	\delta(t_j) = m^{QSO}(t_j) - \left\langle\ S\ \right\rangle(t_j) = -2.5\log\left(\frac{f_{j}^{QSO}}{\sqrt[N]{\Pi_{i}f_{ij}}}\right)
\end{equation}
where $\delta(t_j)$ is the difference in magnitude between the quasar and the average of field stars at time $t_j$ (in flux it reads as the ratio of the object flux with the geometrical average of the field stars). So, as $\left\langle S\right\rangle(t)$ is almost flat, the time series of $\delta(t_j)$ gives us the quasar light curve with respect to a constant standard and then variations of $\delta$ are exclusively atributted to the quasar. To quantify variability we define the paremeter 
\begin{equation}
	{\cal PP}_{var} (\%) = \left(\frac{F_{max}}{F_{min}}-1\right)\cdot 100\%
\end{equation}
 where $F_{max}$ and $F_{min}$ are the maximum and minimum flux in the DLC, then ${\cal PP}_{var}$ is a function of the difference $\delta_{min}-\delta_{max}$ (because $\Delta m \propto \log(F_1/F_2)$). In this way, the ${\cal PP}_{var}$ parameter is just the peak-to-peak variation in flux.\\
A minimum variability is needed in the continuum to detect a correlated variability in the emission lines (which variation is weaker because is averaged in a larger geometry than the continuum source). Previous reverberation campaigns have concluded that this minimum is $\sim$15\%, then, objects with variability larger than that trigger the spectroscopic follow-up. 
%%%%%% SPECTROSCOPIC FOLLOW-UP %%%%%%%%%%%%%%%%%%%%%%%%%%%%%%%%%%%%%%%%%%%%%%%%%%%%%%%%%%%%%%%%%%%%%%%%%%%%%%%%
\subsection{Spectroscopic follow-up.}
As mentioned before, we are triggering the spectroscopy follow up whenever we detect 15\% variability using the 2.5m DuPont Telescope at Las Campanas Obseratory (LCO). We note that at such quasar luminosity, the typical intrinsic continuum variability time scale is 5--10 months. Given the $(1+z)$ factor, in the observed frame the lines are expected to follow the continuum after 1--2 years. During 2005-2006 we obtained spectroscopic observations for the whole sample to establish a zero baseline for future spectroscopy and evaluate the spectroscopic accuracy of our method. In 2007 we have started the actual spectroscopic follow-up of the first quasars. Currently, there are 5 targets which have been followed up since 2007 and there are 3 more quasars reciently (2008) selected for the spectroscopic monitoring.
%%%%%% PRELIMINARY RESULTS %%%%%%%%%%%%%%%%%%%%%%%%%%%%%%%%%%%%%%%%%%%%%%%%%%%%%%%%%%%%%%%%%%%%%%%%%%%%%%%%
\subsection{Preliminary results.}
After 3 years of monitoring we can confirm that, in general, quasars are very slow and low amplitude variables (<0.1 mag). However, our strategy has allowed us to identify a subsample of 8 quasars that present a variability greater than 15\% (about 0.2 magnitudes, shown in Figure 2) which correspond to the 14.2\% of the sample. The variability in some of the objects has been seen to reach up to 20\% and spectroscopic follow up has began for them. On the other hand, the varibility of the field stars $\left\langle\ S\ \right\rangle $ is lower than 3\%. This range is represented in Figure 2 using dot-dashed lines. All points of $\left\langle\ S\ \right\rangle $ fall within this region.\\ Up to date DLC for our targets can be found at the website of the program \small \texttt{http://www.cec.uchile.cl/$\sim$ibotti/reverberation/}.
\normalsize
%\begin{figure}[t!]
%	\centering
%		\includegraphics[width=14cm,height=13cm]{Botti_f2.jpg}
%		\caption{\small DLCs of some of our most variable objects. \it Left: \rm CTQ803 (\emph{a}), CTQ953 (\emph{b}) and CTQ975 (\emph{c}) Notice that the curves not only show monotonic flux variations, but also peaks, crucial features to apply reverberation mapping analysis. \it Right: \rm DLCs of CTQ320 (\emph{a}), J214355 (\emph{b}) and J224743.\\The dot-dashed lines denote 3\% variability amplitude. Field stars average $\left\langle S\right\rangle(t_j)$ lies within this region in all our targets.}
%\end{figure}
\begin{figure}[t!]
\begin{center}
\hbox{
	\includegraphics[width=7.2cm,height=11.4cm]{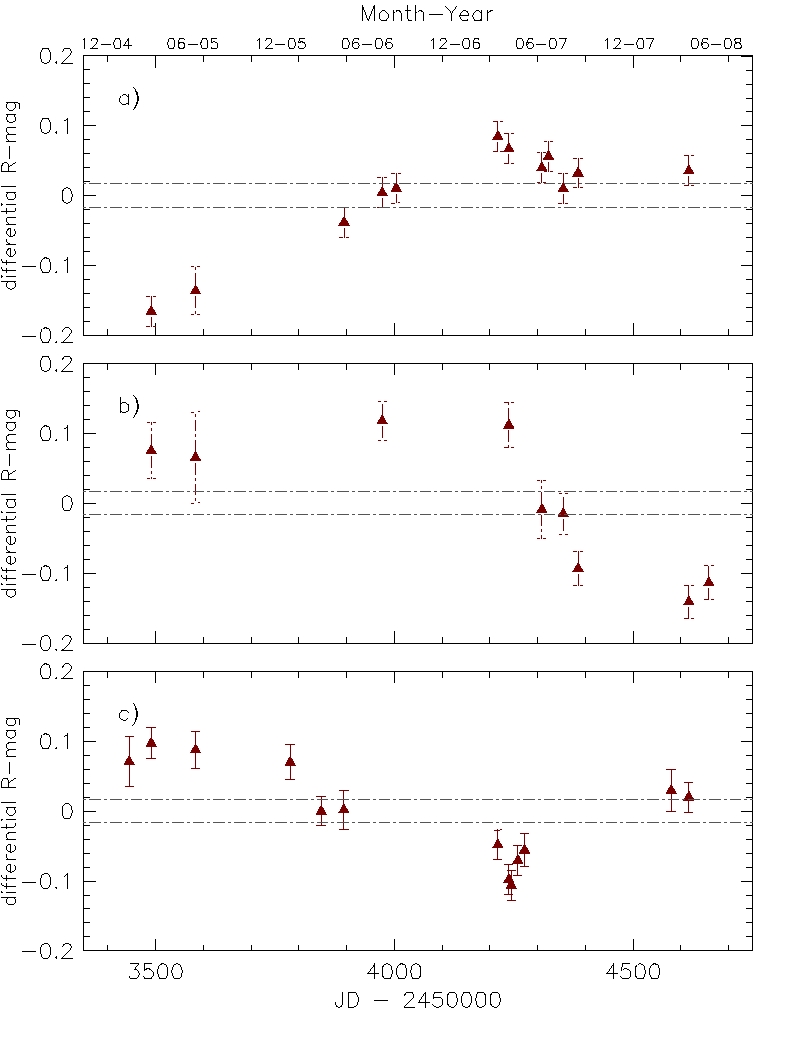}
	\includegraphics[width=7.2cm,height=11.4cm]{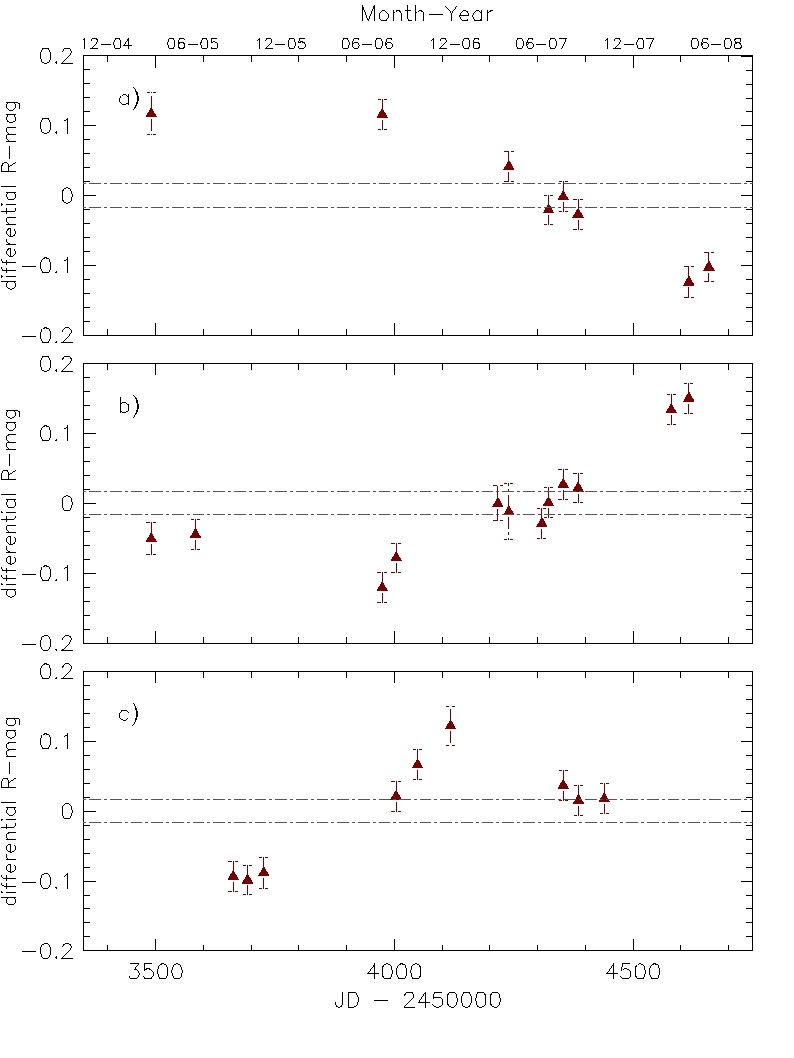}
        }
\caption{\small DLCs of some of our most variable objects. \it Left: \rm CTQ953 (\emph{a}), CTQ975 (\emph{b}) and CTQ320 (\emph{c}). Notice that the curves not only show monotonic flux variations, but also peaks, crucial features to apply reverberation mapping analysis. \it Right: \rm DLCs of J224743 (\emph{a}), J214355 (\emph{b}) and CTQ803 (\emph{c}).}
\end{center}
\end{figure}
%%%%%% SUMMARY AND PERSPECTIVES %%%%%%%%%%%%%%%%%%%%%%%%%%%%%%%%%%%%%%%%%%%%%%%%%%%%%%%%%%%%%%%%%%%%%%%%%%%%%%%%
\section{Summary and Perspectives}
We reviewed the basis of reverberation mapping technique which is a powerful tool to obtain black hole masses in AGN and the Size-Luminosity ($R_{BLR}-L$) relation from which we can estimate the size of the BLR (and then estimate the BH mass) from a single epoch observation. Currently this relation covers almost 5 magnitudes in luminosity, but it is subjected to large uncertainties when is extrapolated to high luminosity AGN. For a better determination of the slope of this relation we need to span the luminosity range of AGN with reverberation mapping results of $10^{40}-10^{48}$ erg/sec. To achieve that aim we present a monitoring campaing of a large sample of high-$z$ quasars which will extend these results by two orders of magnitude, measuring the broad-line region size and black hole mass of luminous AGN at high redshifts. Our strategy has been successful identifying a subsample of high variability ($\ge$15\%) quasars which are being followed spectroscopically since 2007. \\
We expect that in the next 3-6 years reverberation mapping analysis will finally yield the highly anticipated measurement of BH masses in some of the most luminous quasars in our Universe. \\
\acknowledgments 
We acknowledge the financial support from the Fondecyt grant no. 1080603 and Proyecto Fondap de Astrof\'isica 15010003.
%%%%%% REFERENCES %%%%%%%%%%%%%%%%%%%%%%%%%%%%%%%%%%%%%%%%%%%%%%%%%%%%%%%%%%%%%%%%%%%%%%%%%%%%%%%%%%%%%%%%%%%%%%%%

%%%%%%%%%%%%%%%%%%%%%%%%%%%%%%%%%%%%%%%%%%%%%%%%%%%%%%%%%%%%%%%%%%%%%%%%%%%%%%%%%%%%%%%%%%%%%%%%%%%%%%%%%%%%%%%%

\begin{references}%\small
\reference Peterson, B., 2001, in ``The Starburst-AGN Connection'', ed. by I. Arextaga, D. Kunth, R. M\'ujica. p3.
\reference Peterson, B. et al.,  2004, ApJ, \textbf{613}, p682.
\reference Kaspi, S. et al., 2005, ApJ, \textbf{629}, p61.
\reference Bentz, M. et al., 2006, ApJ, \textbf{644} p133.
\end{references}
\end{document}